\documentclass[10pt,letterpaper,twocolumn]{article}
\usepackage{ol2}
\usepackage{hyperref}
\usepackage{amsmath}

\begin{document}

\twocolumn[

\title{Infrared dielectric properties of low-stress silicon nitride }

\vskip -0.15 in

\author{Giuseppe Cataldo,$^{1,2,*}$ James A. Beall,$^3$ Hsiao-Mei Cho,$^3$ \\
Brendan McAndrew,$^1$ Michael D. Niemack,$^3$ and Edward J. Wollack$^1$}

\address{$^1$NASA Goddard Space Flight Center, 8800 Greenbelt Road, Greenbelt, MD 20771, USA\\
$^2$Universities Space Research Association, 10211 Wincopin Circle, Columbia, MD 21044,USA\\
$^3$National Institute of Standard and Technology, 325 Broadway, Boulder, CO 80305, USA\\
$^*$Corresponding author: Giuseppe.Cataldo@nasa.gov}

\vskip -0.10 in

\begin{abstract}
Silicon nitride thin films play an important role in the realization of sensors, filters, and high-performance circuits. Estimates of the dielectric function in the far- and mid-infrared regime are derived from the observed transmittance spectra for a commonly employed low-stress silicon nitride formulation. The experimental, modeling, and numerical methods used to extract the dielectric parameters with an accuracy of approximately 4\% are presented.
\end{abstract}
\vskip -0.20 in
]

\maketitle

The physical properties of silicon nitride thin films, namely low tensile stress, low thermal/electrical conductance, and its overall compatibility with other common materials, have facilitated its use in the micro-fabrication of structures requiring mechanical support, thermal isolation, and low-loss microwave signal propagation (e.g.,~\cite{Goldie, Wang, Martinis, Paik}). Silicon nitride films are amorphous, highly absorbing in the mid-infrared~\cite{Eriksson}, and their general properties are functions of composition~\cite{Taft, Palik}. Here, the optical properties are studied in detail for a membrane with parameters commonly employed in micro-fabrication. 

The silicon nitride optical test films were prepared by a LP-CVD (Low-Pressure Chemical-Vapor-Deposition) process optimized for low tensile stress and refractive index~\cite{Sekimoto}. The 5:1 SiH$_2$Cl$_2$/ NH$_3$ gas ratio employed results in a tensile stress $<100$ MPa and optical index greater than $\sim$ 2~\cite{Makino}. The test structure is shown schematically in Fig. 1 (inset). Double-side-polished silicon (75-mm-diameter, 500-$\mu$m-thick) wafers~\cite{Addison} were used as a mechanically robust handling structure for the SiN$_x$ membranes. A 150-nm thermal oxide was grown on the silicon wafers by wet oxidation at 950$^{\circ}$C for 31 minutes. This layer was subsequently used as an etch stop to protect the nitride during definition of the silicon handling wafer geometry. A low-stress SiN$_x$ layer was then deposited by LP-CVD (e.g., deposition parameters for 2-$\mu$m film are 835$^{\circ}$C for 9.7 hours with pressure 33~Pa and 12~sccm NH$_3$, 59~sccm SiH$_2$Cl$_2$). The wafers were then patterned with a resist mask and SiN$_x$/SiO$_2$ windows formed by deep reactive ion etching which removed all the silicon under the window area. The residual thermal oxide was removed with HF vapor etch leaving a set of uniform SiN$_x$ membranes each with a 10-mm diameter aperture individually suspended from the silicon handling frame. 

The optical tests were performed on SiN$_x$ samples having membrane thicknesses of 0.5 and 2.3~$\mu$m with a uncertainty of 3\%. Fabry-Perot resonators were made by stacking multiple samples with silicon standoff frames between adjacent samples to explore the long-wavelength response of the material in greater detail. The silicon standoffs allowed a vent path for evacuation of air between the nitride membranes. All optical measurements were performed in vacuum with a residual pressure less than 100~Pa.

\begin{figure}[htbp]
	\vskip -0.10 in
	\centering
		\includegraphics[width=0.47\textwidth]{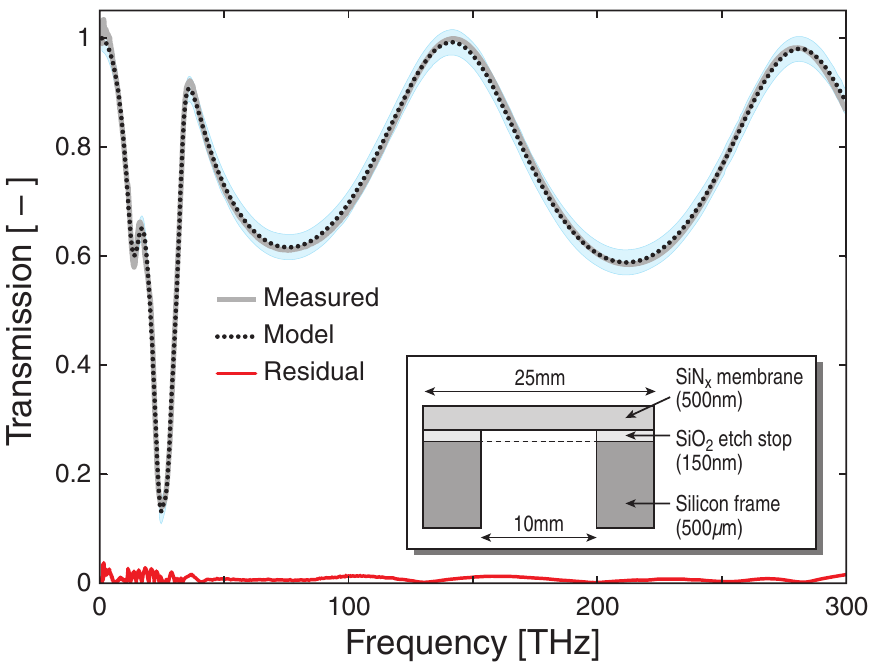}
	\vskip -0.10 in
	\caption{(Color online) Room-temperature transmission of a silicon nitride sample 0.5 $\mu$m thick: measured (grey), model (black dotted), and residual (red). The shaded band's width delimits the estimated 3$\sigma$ measurement uncertainty. A 30~GHz (1~cm$^{-1}$) resolution is employed for the measurement. The insert depicts the geometry of the SiN$_x$ membrane and micro-machined silicon frame.}
	\label{fig:T_500nm}
	\vskip -0.0 in
\end{figure}

\begin{figure}[htbp]
	\vskip -0.25 in
	\centering
		\includegraphics[width=0.45\textwidth]{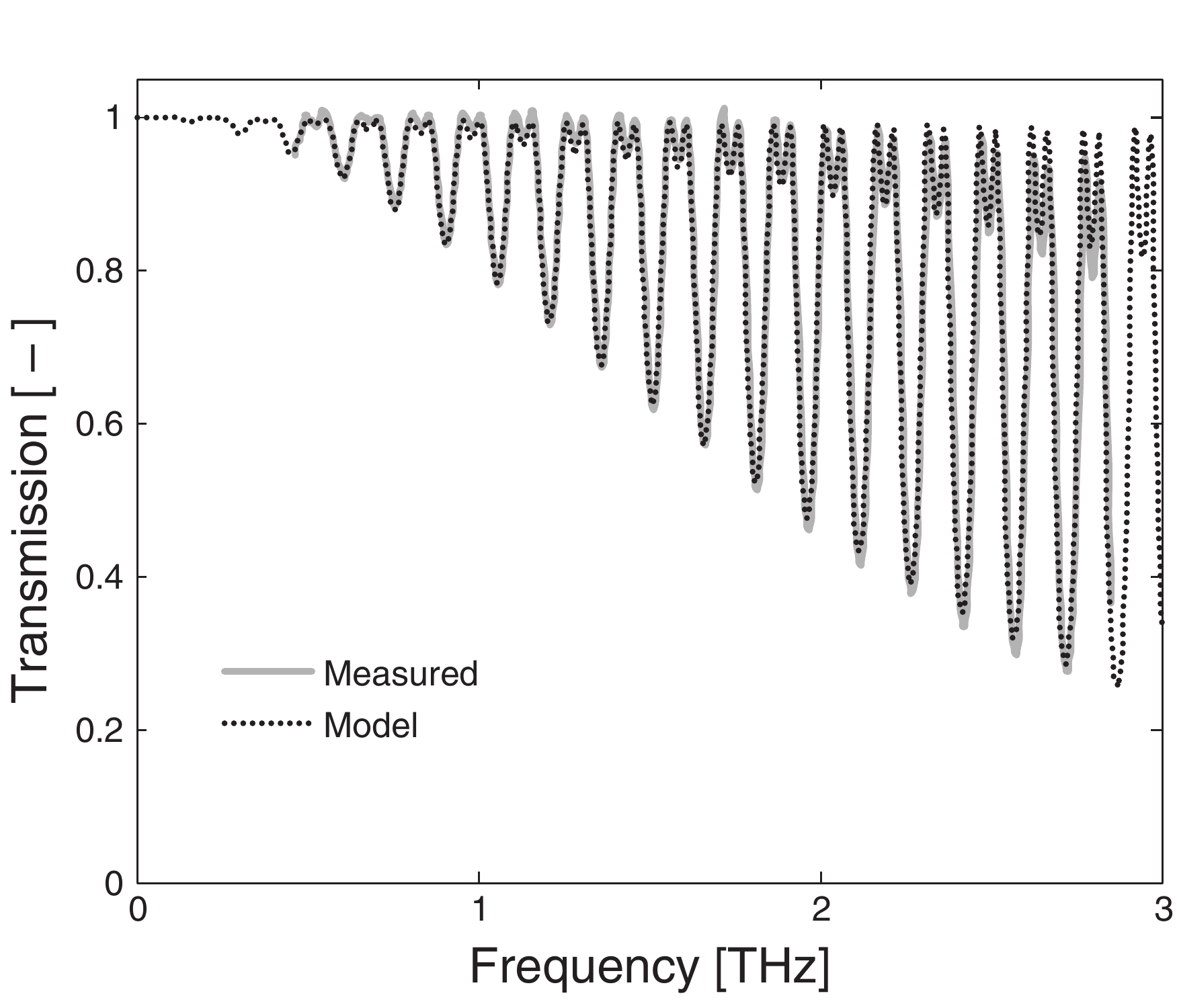}
	\vskip -0.10 in
	\caption{(Color online) Measured (solid grey) and model (black dotted) transmission for a 3-layer stack of silicon nitride samples 2.3~$\mu$m in thickness with 998-$\mu$m intermembrane delays which complements the data shown in Fig.~\ref{fig:T_500nm}. The sample response in the far-infrared was acquired with a resolution of 3~GHz (0.1~cm$^{-1}$).}
	\label{fig:3-layer sample 0-3THz}
	\vskip -0.0 in
\end{figure}

The samples were characterized with a Bruker 125 high-resolution Fourier Transform Spectrometer (FTS) and were measured in transmission at the focal plane of an $f/6$ beam. A number of different sources, beam splitters, and detector configurations were used in combination to provide measurements over the reported spectral range. The single-layer SiN$_x$ sample transmission was measured over an extended range from 15 to 10000~cm$^{-1}$. The mercury lamp and a multilayer Mylar beam splitter were used to access frequencies below 600~cm$^{-1}$.  Additional mid-infrared spectral data up to 2400~cm$^{-1}$ were acquired using a ceramic glow bar source, Ge-coated KBr beam splitter, and room-temperature DTGS detector. The remaining near-infrared data up to 10000~cm$^{-1}$ were taken with a W filament source, Si on CaF$_2$ beam splitter, and a liquid-nitrogen-cooled InSb detector (Fig.~\ref{fig:T_500nm}). Far-infrared data between 15 and 95~cm$^{-1}$ were taken using a mercury arc lamp source and a liquid-helium-cooled 4.2-K bolometer. Mylar beam splitters of 50-, 75- and 125-$\mu$m thicknesses and a multilayer Mylar beam splitter were used during separate scans (Fig.~\ref{fig:3-layer sample 0-3THz}). The resultant  transmission data were merged into a single spectra using a signal-to-noise weighting for subsequent parameter extraction.

The dielectric response is represented as a function of frequency, $\omega$, by the classical Maxwell-Helmholtz-Drude dispersion model \cite{Button}:
\begin{equation}
	\overset{\hat{}}{\varepsilon_r}(\omega)=\overset{\hat{}}{\varepsilon}_\infty+\sum^{M}_{j=1}\frac{\Delta\overset{\hat{}}{\varepsilon}_j\cdot\omega^2_{\mbox{{\tiny\it T}}_j}}{\omega^2_{\mbox{{\tiny\it T}}_j}-\omega^2-i\omega\Gamma'_j(\omega)}
\label{eq:Lorentz}
\end{equation}
where $M$ is the number of oscillators and $\overset{\hat{}}{\varepsilon_r}=\varepsilon'_r+i\varepsilon''_r$ is a complex function of $(5M+2)$ degrees of freedom, which are as follows: the contribution to the relative permittivity $\overset{\hat{}}{\varepsilon}_\infty=\overset{\hat{}}{\varepsilon}_{M+1}$ of higher lying transitions, the difference in relative complex dielectric constant between adjacent oscillators $\Delta\overset{\hat{}}{\varepsilon}_j=\overset{\hat{}}{\varepsilon}_j-\overset{\hat{}}{\varepsilon}_{j+1}$ which serves as a measure of the oscillator strength, the oscillator resonance frequency $\omega_{\mbox{{\tiny\it T}}_j}$, and the effective Lorentzian damping coefficient $\Gamma'_j$, for $j=1,...,M$. The following functional form is used to specify the damping:
\begin{equation}
	\Gamma'_j(\omega)=\Gamma_j\exp{\left[-\alpha_j\left(\frac{\omega^2_{\mbox{{\tiny\it T}}_j}-\omega^2}{\omega\Gamma_j}\right)^2\right]}
\label{eq:Gauss}
\end{equation}
where $\alpha_j$ allows interpolation between Lorentzian $(\alpha_j=0)$ and Gaussian wings $(\alpha_j>0)$ similar to the approach in~\cite{Kim}. The form indicated above enables a more accurate representation of relatively strong oscillator features. 

\begin{figure}[htbp]
	\vskip -0.20 in
	\centering
		\includegraphics[width=0.47\textwidth]{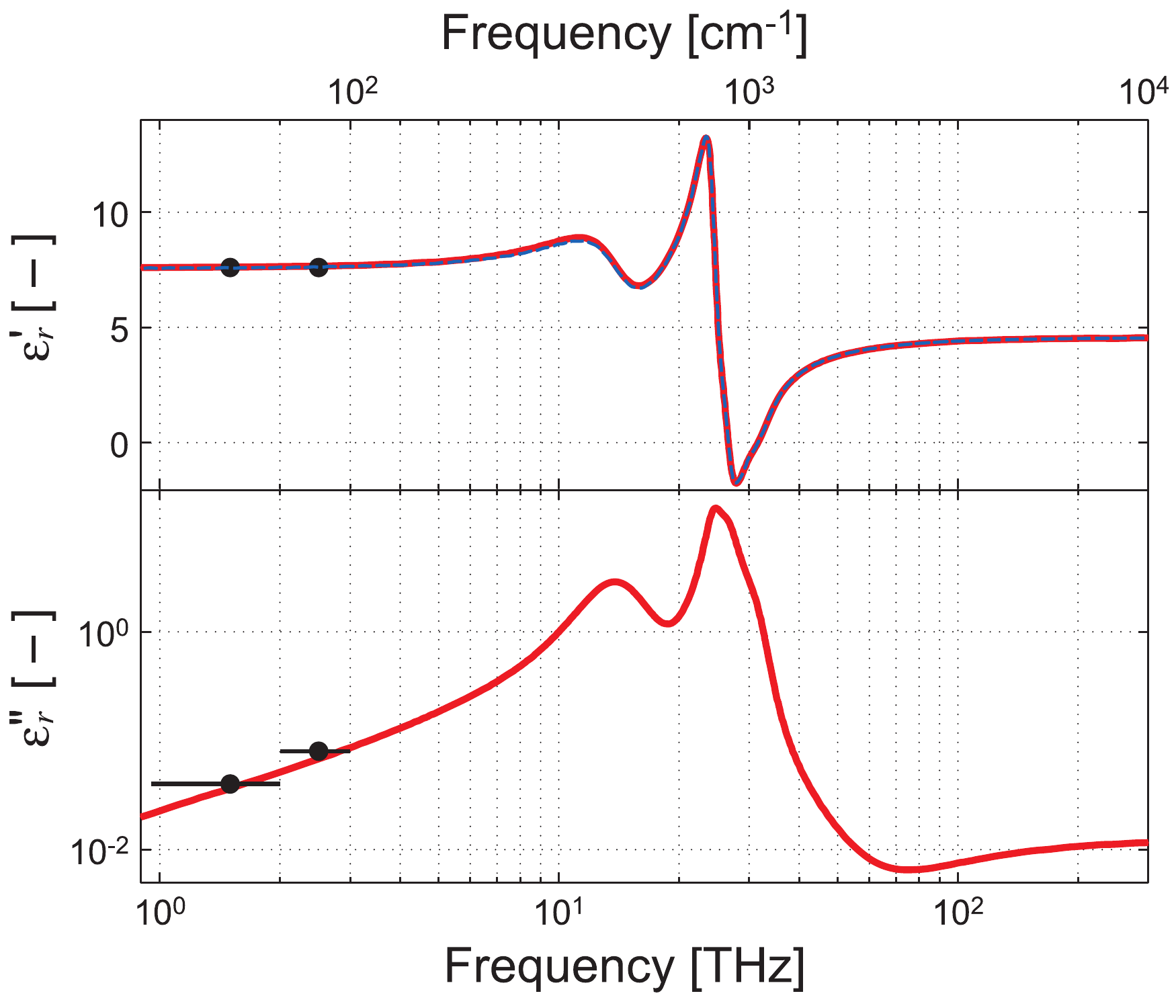}
	\vskip -0.10 in
	\caption{(Color online) Real and imaginary parts (solid red lines) of the dielectric function as extracted from the data shown in Fig.~\ref{fig:T_500nm}. The line thickness is indicative of the propagated $\sim$ 4\% error band. The numerical Hilbert transform of the modeled $\varepsilon_r''(\omega)$ is indicated in the upper panel (dashed blue line) to facilitate comparison with $\varepsilon_r'(\omega)$. The filled symbols indicate the parameters derived from the data presented in Fig.~\ref{fig:3-layer sample 0-3THz}.}
	\label{fig:eps-uncertainty_500nm}
	\vskip -0.20 in
\end{figure}

The impedance contrast between free space and the thin-film sample forms a Fabry-Perot resonator. The observed transmission can be modeled ~\cite{Yeh} as a function of the dielectric response~(Eq. \ref{eq:Lorentz}), thickness, and wavenumber. The dielectric parameters were solved by means of a non-linear least-squares fit of the transmission equation to the laboratory FTS data. Specifically, a sequential quadratic programming (SQP) method with computation of the Jacobian and Hessian matrices~\cite{Biggs, Powell} was implemented. The merit function, $\chi^2$,was used in a constrained minimization over frequency as follows:
\begin{equation}
\underset{\mbox{\footnotesize DOF}}{\mbox{\footnotesize min}}\chi^2=\underset{\mbox{\footnotesize DOF}}{\mbox{\footnotesize min}}\sum^{N}_{k=1}{\left[T(\overset{\hat{}}{\varepsilon_r}(\omega),h)-T_{\mbox{\tiny{FTS}}_k}\right]^2}
\end{equation}
where $N$ is the number of data points, $T$ the modeled transmittance, $T_{\mbox{\tiny FTS}}$ the measured transmittance data, and $h$ is the measured sample thickness. We are guided by the Kramers-Kronig relations in defining constraints for a passive material: $|\overset{\hat{}}{\varepsilon}_j|>|\overset{\hat{}}{\varepsilon}_{j+1}|$, $\varepsilon''_j>0$ and $\overset{\hat{}}{\varepsilon_r}(0)=\overset{\hat{}}{\varepsilon}_1$~\cite{Landau}. For accurate parameter determination the sample should have uniform thickness, be adequately transparent to achieve high signal-to-noise, and have diffuse scattering as a sub-dominate process. The method requires an \textit{a posteriori} numerical verification for Kramers-Kronig consistency. In the example presented here, a numerical Hilbert transform \cite{Mori} of $\varepsilon_r''(\omega)$ reproduces $\varepsilon_r'(\omega)$ to within 2\% (Fig. \ref{fig:eps-uncertainty_500nm}). An alternative method employing reflectivity and phase allows {\it a priori} Kramers-Kronig consistent results~\cite{Nitsche}. However, given the details of the thin-film samples and available instrumentation, this approach was not implemented. 

Figure~\ref{fig:T_500nm} illustrates the measured and modeled results obtained from the analysis of a 0.5-$\mu$m-thick sample. The peak residual in the transmittance is less than 3\% and the $3\sigma=0.023$ uncertainty band indicated corresponds to the 99.7\% confidence level. The standard deviation adopted for the measured data, $\sigma$, was estimated assuming the errors as a function of frequency are uniform and have a reduced $\chi^2$ equal to unity. An additional uncertainty in the FTS normalization influences the dielectric response function at the 1\% level. In addition to the channel spectra, the observed spectrum shows two predominant features at 12 THz and 25 THz. Simulations with $M=2$ oscillators lead to a peak residual on transmission of 5\% and do not enable recovery of the resonance at 25 THz. Using 5 oscillators satisfactorily recovers the observed transmittance and reduces the peak residual by a factor of 4.4. When the resonator's quality factor, $Q_{\mbox{{\scriptsize\it eff}}_j}=\omega_j/\Gamma'_j$, is greater than 5, the data were not reproducible by either a pure Lorentzian oscillator or Eq.~(4.6) in~\cite{Kim}. In these regions, the peak transmission residuals were decreased by a factor $\sim$ 2 through the use of Eq.~(\ref{eq:Gauss}).

In Fig.~\ref{fig:eps-uncertainty_500nm} the values of the real and imaginary components of the dielectric function are illustrated as a function of frequency. The uncertainty in $\overset{\hat{}}{\varepsilon_r}$ was propagated and computed as described in~\cite{NumericalRecipes}. 
Table~\ref{tb:results} contains a summary of the best fit parameters for 5 oscillators, which can be used to reproduce the data shown in Fig.~\ref{fig:eps-uncertainty_500nm}.

\vspace{-0.20 in}
\begin{table}[htbp]
  \centering
  \caption{Fit parameter summary}
    \begin{tabular}{llllll}\\ \hline
	    $j$ & $\varepsilon'_j$ & $\varepsilon''_j$ & $\omega_{\mbox{{\tiny\it T}}_j}/2\pi$ & $\Gamma_j/2\pi$ & $\alpha_j$\\
	    $[-]$ & $[-]$ & $[-]$ & $[$THz$]$ & $[$THz$]$ & $[-]$ \\ \hline
	    1 & 7.582    & 0		       & 13.913   & 5.810   & 0.0001 \\
	    2 & 6.754    & 0.3759      & 15.053   & 6.436   & 0.3427 \\
  	  3 & 6.601    & 0.0041      & 24.521   & 2.751   & 0.0006 \\
      4 & 5.430    & 0.1179      & 26.440   & 3.482   & 0.0002 \\
      5 & 4.601    & 0.2073      & 31.724   & 5.948   & 0.0080 \\
      6 & 4.562    & 0.0124      &          &         &  \\ \hline
    \end{tabular}
  \label{tb:results}
\end{table}	
\vskip -0.05 in

In order to characterize the long-wavelength portion of the dielectric function, Fabry-Perot resonators were realized from 1-, 2-, and 3-layer samples. Representative data for the 3-layer resonator stack is presented in Fig.~\ref{fig:3-layer sample 0-3THz}. A multilayer transfer matrix analysis~\cite{Yeh} is used to extract the dielectric function using the measured SiN$_x$ (2.3 $\mu$m) and silicon spacer (998 $\mu$m) thicknesses. The circular symbols at 1.5 THz and 2.5 THz indicated in Fig.~\ref{fig:eps-uncertainty_500nm} were computed from a composite analysis of the 3 Fabry-Perot measurement sets. The horizontal range indicates the data used in each fit. The best estimates are $\overset{\hat{}}{\varepsilon_r} \approx 7.6+i0.08$ over the range 2-3~THz and $\overset{\hat{}}{\varepsilon_r} \approx 7.6+i0.04$ over 0.4-2~THz. The real component of the static dielectric function derived from the data is in agreement with prior reported parameters for this stoichiometry ~\cite{Paik}. As shown in Fig.~\ref{fig:eps-uncertainty_500nm}, the measurements are internally consistent and represent roughly a factor-of-three reduction in uncertainty relative to prior infrared SiN$_x$ measurements identified by the authors ~\cite{Palik, Taft, Eriksson}. The dielectric parameters reported here are representative of low-stress SiN$_x$ membranes encountered in our fabrication and test efforts.
\vskip -0.08 in

\pagebreak

\section*{Informational Fourth Page}
In this section, please provide full versions of citations to assist reviewers and editors (OL publishes a short form of citations) or any other information that would aid the peer-review process.


\begin{thebibliography}{99}
	
	\bibitem{Goldie}
		D. J. Goldie, A. V. Velichko, D. M. Glowacka, and S. Withington, ``Ultra-low-noise MoCu transition edge sensors for space applications,'' Appl. Phys. \textbf{109}, 084507 (2011).
 	
	\bibitem{Wang}
		G. Wang, V. Yefremenko, V. Novosad, A. Datesman, J. Pearson, R. Divan, C. L. Chang, L. Bleem, A. T. Crites, J. Mehl, T. Natoli, J. McMahon, J. Sayre, J. Ruhl, S. S. Meyer, and J. E. Carlstrom, ``Thermal Properties of Silicon Nitride Beams Below One Kelvin,'' IEEE Trans. Appl. Superconductivity \textbf{21} (3), 232--235 (2011).
	
	\bibitem{Martinis}
		J. M. Martinis, K. B. Cooper, R. McDermott, M. Steffen, M. Ansmann, K. D. Osborn, K. Cicak, S. Oh, D. P. Pappas, R. W. Simmonds, and C. C. Yu, ``Decoherence in Josephson Qubits from Dielectric Loss'', Phys. Rev. Lett. \textbf{95} (21), 210503 (2005).
	
	\bibitem{Paik}
		H. Paik and K. D. Osborn, ``Reducing Quantum-Regime Dielectric Loss of Silicon Nitride for Superconducting Quantum Circuits,'' Appl. Phys. Lett. \textbf{96} (7), 072505 (2010).
	
	\bibitem{Eriksson}
		T. Eriksson, S. Jiang, and C. Granqvist, ``Dielectric function of sputter-deposited silicon dioxide and silicon nitride films in the thermal infrared,'' Appl. Opt.  \textbf{24,} 745--746 (1985).
	
	\bibitem{Taft}
		E. A. Taft, ``Characterization of Silicon Nitride Films,'' J. Electrochem. Soc. \textbf{118,} 1341--1346 (1971).
		
	\bibitem{Palik}		
		E. D. Palik, \emph{Handbook of Optical Constants of Solids}, Vol. 1 (Elsevier, 1998), pp.~771--774. 

 	\bibitem{Sekimoto}
 		M. Sekimoto, H. Yoshihara, and T. Ohkubo, ``Silicon Nitride Single-Layer X-Ray Mask,'' J. Vac. Sci. Technol. \textbf{21} (4), 1017--1021 (1982).
 
	\bibitem{Makino}
		T. Makino, ``Composition and Structure Control by Source Gas Ratio in LPCVD SiN$_x$,'' J. Electrochem. Soc. \textbf{130} (2), 450--455 (1983).

	\bibitem{Addison}
 		Addison Engineering, 150 Nortech Parkway, San Jose, CA 95134. (Orientation $<$100$>$, Czochralski, p-type B doped, bulk resistivity $<$~0.005~$\Omega\cdot$cm)
		
	\bibitem{Button}
		F. Gervais, ``High-Temperature Infrared Reflectivity Spectroscopy by Scanning Interferometry'' in \emph{Electromagnetic Waves in Matter}, Part I, Vol. 8 (Infrared and Millimeter Waves), K. J. Button, eds. (Academic Press, London, 1983), pp. 284--287.

	\bibitem{Kim}
		C. C. Kim, J. W. Garland, H. Abad, and P. M. Raccah, ``Modeling the optical dielectric function of semiconductors: Extension of the critical-point parabolic-band approximation,'' Phys. Rev. B \textbf{45} (20), 11749 (1992).

	\bibitem{Yeh}
		P. Yeh, \emph{Optical Waves in Layered Media} (Wiley, New York, 1988), pp.~102--111.

	\bibitem{Biggs}
		M. C. Biggs, ``Constrained Minimization Using Recursive Quadratic  Programming,'' in \emph{Towards Global Optimization}, L. C. W. Dixon and G. P. Szergo, eds. (North-Holland, 1975), pp.~341--349.

	\bibitem{Powell}
		M. J. D. Powell, ``Variable Metric Methods for Constrained Optimization,'' in \emph{Mathematical Programming: The State of the Art}, A. Bachem, M. Grotschel  and B. Korte, eds. (Springer Verlag, 1983), pp.~288--311.

	\bibitem{Landau}
		L. D. Landau and E. M. Lifshitz, \emph{Electrodyamics of Continuous Media}, Vol. 8 (Pergamon Press, 1960), pp.~253--262.

	\bibitem{Mori}
		M. Mori and T. Ooura, ``Double Exponential Formulas for Fourier Type Integrals with a Divergent Integrand,'' in \emph{Applicable Analysis}, Vol. 2 (World Scientific Series, 1993), pp.~301--308.

	\bibitem{Nitsche}
		R. Nitsche and T. Fritz, ``Determination of model-free Kramers-Kronig consistent optical constants of thin absorbing films from just one spectral measurement: Application to organic semiconductors,'' Phys. Rev. B \textbf{70} (19), 195432 (2004).

	\bibitem{NumericalRecipes}
		W. H. Press, S. A. Teukolsky, W. T. Vetterling, and B. P. Flannery, \emph{Numerical Recipes - The Art of Scientific Computing} (Cambridge University Press, 2007), pp.~799--806.
		
\end{thebibliography}
\end{document}